\documentclass[aps,onecolumn,11pt]{revtex4}

\usepackage[dvips]{graphicx} 

\begin{document}
\draft

\title{Current distribution and ac loss for a superconducting rectangular strip with in phase alternating current and applied field}
\author{E. Pardo$^{\rm a,b}$, F. G\"om\"ory$^{\rm a}$, J. \v Souc$^{\rm a}$, J. M. Ceballos$^{\rm a,c}$}
\address{$^{\rm a}$ Institute of Electrical Engineering, Slovak Academy of Sciences, 841 04 Bratislava, Slovak Republic.\\
$^{\rm b}$ Grup d'Electromagnetisme,
Departament de F\'\i sica, Universitat Aut\`onoma Barcelona,
08193 Bellaterra (Barcelona), Catalonia, Spain.\\
$^{\rm c}$ Laboratorio Benito Mahedero de Aplicaciones Eléctricas de los Superconductores, Escuela de Ingenierías Industriales,  Universidad de Extremadura. Apdo 382. Avda de Elvas s/n. 06071 Badajoz, Spain.}

\begin{abstract}

The case of ac transport at in-phase alternating applied magnetic fields for a superconducting rectangular strip with finite thickness has been investigated. The applied magnetic field is considered perpendicular to the current flow. We present numerical calculations assuming the critical state model of the current distribution and ac loss for various values of aspect ratio, transport current and applied field amplitude. A rich phenomenology is obtained due to the metastable nature of the critical state. We perform a detailed comparison with the analytical limits and we discuss their applicability for the actual geometry of superconducting conductors. We also define a loss factor which allow a more detailed analysis of the ac behavior than the ac loss. Finally, we compare the calculations with experiments, showing a significant qualitative and quantitative agreement without any fitting parameter.

\end{abstract}

\maketitle

\section{Introduction}

The behavior of a superconductor transporting an alternating current or exposed to a magnetic field varying in time have been a wide subject of study from the early 1960's \cite{bean62PRL,london63PhL,wilson,carr}.
However, the case of a simultaneous alternating transport current and applied magnetic field remains unclear. This situation is found in superconductor windings where each turn feels the magnetic field of all the other. Windings are present in many applications, like ac magnets, transformers and motors \cite{hull03RPP,oomen03SST,oomen04SST,larbalestier01Nat}. From the practical point of view, it is of fundamental importance to understand, predict and, eventually, reduce the energy loss (or ac loss) in the superconductor. Actually, the reduction of the ac loss is vital for the applicability of superconductor electrical technology \cite{oomen04SST,larbalestier01Nat}. 
The study of the ac loss is also interesting for material science, since in can be used for characterizing superconducting samples \cite{gomory97SST,coatedsingle,gherardi97SST,miyagi02IES}.

The superconductors suitable for electrical applications are hard type II ones \cite{bean64RMP}. Nowadays there is a great scientific effort in the development of silver sheathed ${\rm Bi_2Sr_2Ca_2Cu_3O_{10}}$ (Ag/Bi-2223) tapes and YBa$_2$Cu$_3$O$_{7-\delta}$ (YBCO) coated conductors, which are high-temperature superconductors, and MgB$_2$ wires \cite{hull03RPP,larbalestier01Nat}. These superconducting tapes and wires have a cross-section roughly rectangular or elliptical. In this work we will consider wires with rectangular cross-section (or rectangular bars), leaving those with elliptical cross-section for further studies. We also restrict to the situation when the ac applied field is uniform and in phase with the transport current.

Hard type II superconductors can be well described by the critical-
state model (CSM) proposed by Bean and London
\cite{bean62PRL,london63PhL}, which assumes that the magnitude of the
local current density cannot be higher than a certain critical value
$J_{\rm c}$.

For the situation of only transport current, the CSM was first applied by London and Hancox in the early 1960's in order to analytically describe simple geometries, like infinite cylinders and slabs, \cite{london63PhL,hancox66IEP}. Later, an important step forward was done by Norris, who analytically deduced the current distribution and the ac loss for an infinitely thin strip by means of conformal mapping transformations \cite{norris70JPD}. The case of a strip with finite thickness can only be solved numerically, as done by Norris \cite{norris71JPD}, Fukunaga {\it et al} \cite{fukunaga98APL}, D\"aumling \cite{daumling98SST} and Pardo {\it et al} \cite{aclosrec}.

The first analysis about the CSM with only ac applied field was done by Bean for an slab with applied field parallel to the surface \cite{bean64RMP}. The case of a thin strip with perpendicular applied field was analytically solved by Brandt {\it et al} \cite{brandt93EPL} following the Norris' technique \cite{norris70JPD}. The current distribution for a strip with finite thickness were numerically calculated by Brandt \cite{brandt96PRB} and Prigozhin \cite{prigozhin96JCP}, and the ac loss by Pardo {\it et al} \cite{acxrec}. 

Concerning the case of simultaneous alternating transport current and applied field, the most significant published calculations within the CSM are the following. In late 1970's Carr analytically derived the ac loss for an infinite slab in parallel applied field \cite{carr79IEM}. In the 1990's Brandt and Zeldov {\it et al} analytically calculated the current distribution in a thin strip using conformal transformations  for the situation that the transport current and the applied field increase monotonically \cite{brandt93PRB,zeldov94PRB}. Although these works provide different formulae, they are actually equivalent \footnote{It can be seen after doing some algebra using that $\arcsin (ix)=i\, {\rm arcsinh}(x)$, $\arctan (ix)=i\, {\rm arctanh}(x)$ and the definition of ${\rm arcsinh}$ and ${\rm arctanh}$.}. Moreover, Brandt studies the values of transport current and applied field for which these formulae are valid, obtaining that they are not applicable for high fields and low currents \cite{brandt93PRB}. From that current distribution, Sch\"onborg analytically calculated the ac loss for a thin strip \cite{schonborg01JAP}. 

For strips with finite thickness, there are no published works dealing with the simultaneous application of alternating currents and magnetic fields in a superconductor in the CSM, according to our knowledge. However, there are several theoretical works assuming a relation between the electrical field ${\bf E}$ and the current density ${\bf J}$ as ${\bf E}({\bf J})=E_c(|{\bf J}|/J_c)^n{\bf J}/|{\bf J}|$, where $E_c$ is an arbitrary value and $n$ is a positive exponent. The current distribution and the ac loss  are calculated in Refs. \cite{amemiya98bPhC,yazawa98PhC,stavrev02bIES,tonsho03IES,enomoto04PhC} and Refs. \cite{yazawa98PhC,amemiya98bPhC,zannella01IES,tonsho03IES}, respectively, for several values of the alternating transport current and applied field. These published results are incomplete, not covering the whole range of combinations of ac current and ac field. This contrasts with the extensive experimental study that has been done for Ag/Bi-2223 tapes \cite{rabbers99IES,magnusson99IES,ashworth99aPhC,ashworth00PhC,tonsho03IES,amemiya03aPhC} and YBCO coated conductors \cite{ashworth00JAP,ogawa03IES}.

The objective of this paper is to provide a systematic study of the current distribution and ac loss for a superconducting strip of finite thickness assuming the CSM under simultaneous application of an alternating transport current and field. The effect of mainly three factors are considered: the aspect ratio of the cross-section and the aplitudes of the transport current the applied magnetic field. We also study the applicability of the CSM to actual superconducting tapes and wires of such shape.

This paper is structured as follows. In Sec. \ref{s.mummet} we present the numerical method used for the calculations and we discuss some general features, mainly about the energy of the system. The results and their discussion are presented in Sec. \ref{s.resdis}. In Sec. \ref{s.exp} the comparison with data measured in a high-temperature superconducting tape is reported. Finally, in Sec. \ref{s.concl} we present our conclusions.


\section{Numerical method and general considerations}
\label{s.mummet}

Let us consider an infinitely long superconductor along the $z$ axis with rectangular cross-section with dimensions $2a\times 2b$ in the $x$ and $y$ directions, respectively, Fig.\ \ref{f.sketch}. The origin of coordinates is taken in the center of the strip. We study here the situation that the superconductor carries a sinusoidal time-varying current $I(t)=I_{\rm m}\cos{\omega t}$ simultaneously immersed in a uniform in-phase ac applied field ${\bf H}_{\rm a}(t)={\bf H}_{\rm m}\cos{\omega t}$ oriented in the $y$ direction. It is shown below (Secs.\ \ref{s.Jcalc} and \ref{s.Qcalc} ) that our results are not only independent on $\omega$ but also on the specific time waveform of $I$ and $H_a$, similarly to the case of only transport current or magnetic field \cite{bean64RMP,london63PhL}.

In our calculations we will consider that first $I$ and ${\bf H}_{\rm a}$ are increased from zero to their maximum, starting from the zero-field cooled state of the superconductor. We call this process the initial stage. Following this stage, we regard the reversal one for which the current and applied field are decreased from ${\bf H}_{\rm m}$ and $I_{\rm m}$, respectively, to $-{\bf H}_{\rm m}$ and $-I_{\rm m}$. Next, the applied field and current are increased back to their maximum, closing the ac cycle. We refer to this latter stage as the returning one.

\subsection{The critical-state model in strips}
\label{s.csm}

We assume that the superconductor obeys the CSM with  constant critical-current density $J_{\rm c}$ \cite{bean62PRL}. The CSM corresponds to supposing a multivalued relation of electrical field $\bf E$ against current density $\bf J$ such that ${\bf E}=E(|{\bf J}|) {\bf J}/|{\bf J}|$ with an $E(|{\bf J}|)$ that only takes finite values for $|J|=J_{\rm c}$, being zero for $|J|<J_{\rm c}$ and infinity for $|J|>J_{\rm c}$ \cite{prigozhin97IES}. For an infinitely long strip along the $z$ direction, the current density and the electrical field inside the superconductor are also in the $z$ direction and they can be considered as the scalar quantities $J$ and $E$, respectively. Although in principle $|J|$ in the CSM can be lower than $J_{\rm c}$, in a superconducting strip $J$ only takes the values 0 or $\pm J_{\rm c}$ \cite{badia02PRB}. 

Let us start with the introduction of the main features of the current distribution in the initial stage for the case of transport current only (i.e. $H_{\rm a}=0$) or when solely the magnetic field is applied ($I=0$).

The behavior of a superconducting strip in the critical state model with $H_{\rm a}=0$ and uniform $J_{\rm c}$ is detailed in \cite{norris70JPD,carr04bPhC,aclosrec}. In the initial stage, for any $I>0$ lower than the critical current, $I_{\rm c}=4abJ_{\rm c}$, there exist a zone with $J=0$ surrounded by another one with $J=J_{\rm c}$. The region with $J=0$ is usually called the current-free core. In this zone the electrical field is zero because in the CSM $E(J=0)=0$. With increasing $I$, the region with $J=J_{\rm c}$ monotonically penetrates from the whole surface inwards and the current-free core shrinks, until it disappears when $I$ reaches $I_{\rm c}$. In the CSM, $I$ cannot overcome $I_{\rm c}$ since it is assumed that $|J|\le J_{\rm c}$. 

The situation when a magnetic field is applied to a superconducting strip that is not transporting any net current is described in \cite{brandt96PRB,prigozhin96JCP}. For this case, the current distribution is antisymmetric to the $yz$ plane. In the initial stage with $H_{\rm a}>0$, there are a zone with $J=J_{\rm c}$ in the right half and another one with $J=-J_{\rm c}$ in the left half expanding from the surface to a current-free core between them. Throughout this paper, we call the border between regions with different $J$ as a current front. It is important to notice that in the current fronts $J$ vanishes and, then, so does $E$. With increasing $H_{\rm a}$, the cross-section of the current-free core shrinks until it becomes a point at $(x,y)=(0,0)$ at the characteristic field $H_{\rm p}$, that is called the penetration field. At fields higher than $H_{\rm p}$, the current distribution is the same as for $H_{\rm a}=H_{\rm p}$. 

When we now consider the simultaneous action of an applied magnetic field on a conductor transporting a nonzero current, one can expect that the qualitative behavior of the current distribution is similar to that one for only transport current or applied field. However, for some situations of $I$ and $H_{\rm a}$ the current distribution presents a different behavior. We discuss this aspect in more detail below (Sec. \ref{s.prof}).

\subsection{Minimization principle for the critical state model}
\label{s.minF}

As discussed by several authors, such as Prigozhin \cite{prigozhin96JCP,prigozhin97IES}, Badia and Lopez \cite{badia01PRL,badia02PRB}, Bhagwat {\it et al} \cite{bhagwat94PhC}, and Sanchez {\it et al} \cite{sanchez01PRB,tapesfull,tranarr}, the distribution of current density for a superconductor assuming the critical state model is such that it minimizes a certain functional. The functionals introduced in \cite{prigozhin96JCP,prigozhin97IES,badia01PRL,bhagwat94PhC} are equivalent, while in \cite{sanchez01PRB} it is proposed the magnetic energy as the quantity to be minimized. As shown in \cite{prigozhin96JCP,prigozhin97IES}, the principle of minimization of the functional, $\cal F$, can be derived from fundamental considerations. In Sec. \ref{s.FW} we demonstrate that  the minimization of $\cal F$ is equivalent to minimizing the magnetic energy provided that in the initial stage the current front  penetrates monotonically from the surface inwards. Some of the situations presented in this paper do not satisfy this condition; therefore we use the minimization of $\cal F$ as follows.

Let consider the case of an infinitely long superconductor extended along the $z$ direction carrying a transport current $I$ and immersed in a uniform applied field in the $y$ direction $H_{\rm a}$, Fig.\ \ref{f.sketch}. With this geometry, the current density is in the $z$ direction and, therefore, so is the vector potential $\bf A$ if we assume the gauge $\nabla\cdot {\bf A}=0$. Then, we can regard these quantities as scalar. Following the notation of Prigozhin \cite{prigozhin97IES}, the current at a certain time distributes in such a way that it minimizes the functional
\begin{eqnarray}
\label{Fdef}
{\cal F}[J] & = & \frac{1}{2}\int_S J({\bf r}) A_J({\bf r}){\rm
d}S-\int_S J({\bf r}) {\hat A}_J({\bf r}){\rm
d}S \nonumber\\
& + & \int_S J({\bf r})[A_{\rm a}({\bf r})-{\hat A_{\rm a}}({\bf r})] {\rm d}S,
\end{eqnarray}
with the constrains
\begin{eqnarray}
I & = & \int_S J({\bf r}){\rm d}S \label{consI}\\
|J| & \leq & J_{\rm c} \label{consJc},
\end{eqnarray}
where $S$ is the superconductor cross-section, $A_J$ is the vector potential created by $J$, $A_{\rm a}$ is the vector potential from the external field, and the quantities with hat correspond to those at the previous time layer. For infinitely long geometry, $A_J$ can be calculated from
\begin{equation}
\label{Adef}
A_{J}({\bf r})=-\frac{\mu_0}{4\pi}\int_{S}J({\bf
r'})\ln\left[(y-y')^2+(x-x')^2\right]{\rm d}S'.
\end{equation}
Defining the current density variation $\delta J\equiv J-{\hat J}$, we
obtain from Eqs. (\ref{Fdef}) and (\ref{Adef}) that the current
density which minimizes the functional $\cal F$ also minimizes the
functional ${\cal F}'$, defined as
\begin{equation}
\label{F'def}
{\cal F}'[\delta J]\equiv\frac{1}{2}\int_S \delta J({\bf r}) \delta
A_J({\bf r}){\rm d}S + \int_S \delta J({\bf r})\delta A_{\rm a}({\bf r}) {\rm d}S,
\end{equation}
where $\delta A_J$ is the vector potential created by $\delta J$ and $\delta A_{\rm a}\equiv A_{\rm a}-{\hat A_{\rm a}}$.


\subsection{Minimization of $\cal F$ and magnetic energy}
\label{s.FW}

In this section we demonstrate that, for the initial stage, the principle of minimization of $\cal F$ is equivalent to the magnetic energy minimization (MEM) provided that current density penetrates monotonically from the surface inwards and the rate of increasing the transport current is proportional to the rate of increasing the applied field.

For calculating the magnetic energy, we assume that the transport current in the strip of Fig.\ \ref{f.sketch} returns through another identical one at a large distance $D$ ($D\gg a,b$). In the following, we consider that the returning strip is centered at $(x,y)=(D,0)$ \cite{tranarr}. Using the general formula for the magnetic energy in an infinitely long circuit $W=(1/2)\int_{S_{xy}}J({\bf r})A_J({\bf r})+\int_{S_{xy}}J({\bf r})A_{\rm a}({\bf r})$, where $S_{xy}$ refers to the whole $xy$ plane area, we obtain that the magnetic energy per strip, $W'$, is 
\begin{equation}
\label{W'def}
W'=\frac{1}{2}\int_{S}J({\bf r})A_J({\bf r})+\int_{S}J({\bf r})A_{\rm a}({\bf r}),
\end{equation}
ignoring constant terms that are irrelevant for MEM. $W'$ of Eq.\ (\ref{W'def}) is independent of the position of the returning strip, as long as it is placed at a enough large distance from the other strip.

In order to compare the minimization of $\cal F'$ at every time layer with the magnetic energy minimization for a $J$ in the initial stage, we do the following \cite{tranarr}. Given a physical $J$, we divide it into $n$ terms $\delta J_i$, so that $J({\bf r})=\sum_{i=1}^{n} \delta J_i({\bf r})$, being $n$ a large number. We choose these terms as the actual current density increments in the initial stage corresponding to the time layer at $t=t_i$, $\delta J_i=\delta J(t=t_i)$ with $t_i>t_{i-1}$. If the current front penetrates monotonically from the surface inwards with increasing $I$, each $\delta J_i$ encloses a current-free core. Furthermore, if $n$ is very high, $\delta J_i$ is nonzero in a thin layer only, so that the vector potential variation at time $t_i$, $\delta A_i\equiv \delta A_{J,i}+\delta A_{{\rm a},i}$, is almost uniform in the layer and
\begin{equation}
\label{dAlayer}
{\cal F'}[\delta J_i]\approx \frac{1}{2}\delta A_i^c (I_i-I_{i-1})+\frac{1}{2}\int_S \delta J_i \delta A_{{\rm a},i},
\end{equation}
where $\delta A_i^c$ is the variation of the vector potential in the current-free core and $I_i$ is the transport current at time $t_i$ with $I_{i=0}=0$.

From Eqs. (\ref{W'def}), (\ref{Adef}) and (\ref{dAlayer}), we can separate $W'$ as
\begin{eqnarray}
\label{W'sep}
W' & = & \frac{1}{2} \int_S J'A_{J}'{\rm d}S+\int_S J'A_{\rm a}'{\rm d}S\nonumber\\
& + & \delta A_1^c \left(I-I_{1}/2\right) \nonumber\\
& + & \frac{1}{2}\int_S \delta J_1 \left(A_{\rm a}-A_{{\rm a},1}/2\right) {\rm d}S,
\end{eqnarray}
where the label `1' corresponds to the first increment of current density set in the superconductor, and $J'\equiv J-\delta J_1$, $A'\equiv A_J'-\delta A_{J,1}$ and $A_{\rm a}'\equiv A_{\rm a}-\delta A_{{\rm a},1}$. To obtain Eq.\ (\ref{W'sep}) we used that all the current density created after $\delta J_1$ lies inside the current-free core of $\delta J_1$.

Following the same steps for all $\delta J_i$, we find that
\begin{eqnarray}
\label{W'Aic}
W' & \approx & \sum_{i=1}^{n} \bigg[\frac{1}{2}\delta A_i^c(2I-I_i-I_{i-1}) \nonumber\\
&  & + \frac{1}{2}\int_S \delta J_i (2A_{\rm a}-A_{{\rm a},i}-A_{{\rm a},i-1}){\rm d}S\bigg]
\end{eqnarray}
If the final current density $J$ is reached by increasing $I$ and $A_{\rm a}$ monotonically from zero and the increase rates of $I$ and $A_{\rm a}$ are proportional to each other, then $2A_{\rm a}-A_{{\rm a},i}-A_{{\rm a},i-1}=\delta A_{{\rm a},i}(2I-I_i-I_{i-1})/(I_i-I_{i-1})$. Inserting this into Eq.\ (\ref{W'Aic}) and using Eq.\ (\ref{dAlayer}), we obtain
\begin{equation}
\label{WF}
W'=\sum_{i=1}^{n}{\cal F}'[\delta J_i]\left(\frac{2I-I_i-I_{i-1}}{I_i-I_{i-1}}\right).
\end{equation}

From Eq.\ (\ref{WF}) we directly deduce that when minimizing ${\cal F'}[\delta J_i]$ for each time layer, $W'$ is also minimized, since $I$ and $I_i$ are fixed external parameters. If the $\delta J_i$ which minimizes ${\cal F'}[\delta J_i]$ is unique, the $J=\sum_{i=1}^{n} \delta J_i$ minimizing $W'$ is also unique.


\subsection{Calculation of current distribution}
\label{s.Jcalc}

We calculate the current distribution by minimization of ${\cal F}'[\delta J]$ of Eq.\ (\ref{F'def}) for each time layer as follows.

As done in \cite{aclosrec,acxrec,tranarr}, each superconducting strip is divided into $N=2n_x\times 2n_y$ elements with dimensions $a/n_x \times b/n_y$; current density is assumed to be uniform in each element. In order to obtain a smoother current front, we allow the current density to have discrete values below $J_{\rm c}$, that is, $J=kJ_{\rm c}/m$ with $k$ being an integer number from 1 to a maximum value $m$. As discussed in Refs. \cite{aclosrec,tranarr}, this reduces the discretization error in our ac loss calculations. In this paper we use between $N=12000$ and $16000$ elements and $m=20$ current steps. 

Given a current $\hat I$, applied vector potential $\hat A_{\rm a}$ and a current density $\hat J$, we calculate the current density variation $\delta J$ if the current is changed into $I$ and the applied vector potential into $A_{\rm a}$, as follows.

First, we find the element with $sJ<J_{\rm c}$, being $s={\rm sign}({I-\hat I})$, where increasing the current by $\Delta I=sJ_{\rm c}ab/n_xn_ym$ produces the minimum increase of ${\cal F}'$ and we repeat the procedure until the current reaches $I$. Then, we find the elements where changing the current by $\Delta I$ and $-\Delta I$, respectively, reduces the most ${\cal F}'$ and $|J|$ does not exceed $J_{\rm c}$; we repeat this procedure until varying the current in any pair of elements will increase $\cal F'$ instead of lowering it. This allows the creation of regions with current density opposite to $I$. Setting the current in this way, we ensure that the constrains of Eqs. (\ref{consI}) and (\ref{consJc}) are fulfilled. In this procedure the time do not play any role, so that the resulting current distribution is independent on the specific current (and field) waveform.

The variation of ${\cal F}'$, $\Delta{\cal F}'$, due to a variation of current $\Delta I$ in the element $j$ can be calculated from Eq.\ (\ref{F'def}) and $A_{\rm a}=-\mu_0H_{\rm a} x$ taking into account the division into elements of the tape, with the result
\begin{eqnarray}
\label{F'num}
\Delta{\cal F}'_j & = & \sum_{k=1}^n \delta I_k\Delta I C_{jk} +\frac{1}{2} (\Delta I)^2C_{jj} \nonumber\\
& - & \mu_0(H_{\rm a}-{\hat H}_{\rm a})\Delta Ix_j,
\end{eqnarray}
where $\delta I_k$ is the current flowing through the element $k$ induced after the change of $I$ and $H_{\rm a}$, $x_j$ is the $x$ coordinate of the center of element $i$, and $C_{jk}$ are geometrical parameters calculated in the appendix of \cite{tranarr}.

For simplicity, we consider a constant variation of $I$ (and $H_{\rm a}$) between different time layers inside each half cycle. In this paper we use between 80 and 320 time layers.

It can be demonstrated that the current distribution in the reverse and returning stages can be obtained from that one in the initial stage as long as in that stage the current front penetrates monotonically from the surface inwards with increasing $I$ (and $H_{\rm a}$) Ref. \cite{tranarr}. Specifically, the current distribution in the reverse and returning stages, $J_{\rm rev}$ and $J_{\rm ret}$, are, respectively,
\begin{eqnarray}
J_{\rm rev}(I) & = & J_{\rm in}(I_{\rm m})-2J_{\rm in}[(I_{\rm m}-I)/2], \label{JrevJin}\\
J_{\rm ret}(I) & = & - J_{\rm in}(I_{\rm m})+2J_{\rm in}[(I_{\rm m}+I)/2], \label{JretJin}
\end{eqnarray}
where $J_{\rm in}$ is the current distribution in the initial stage \footnote{Unfortunately, we noticed that in Sec. IIC of Ref. \cite{tranarr} there is a typing error in the equation for $J_{\rm ret}$.}.


\subsection{Calculation of ac loss}
\label{s.Qcalc}

The power loss per unit volume in a conductor is ${\bf J}\cdot{\bf E}$. Then, the ac loss per unit length and cycle $Q$ in the superconducting strip is
\begin{equation}
\label{Q}
Q=\oint {\rm d}t \int_S J(x,y;t)E(x,y;t){\rm d}x{\rm d}y,
\end{equation}
where the time integral is over one period. 

Since the electrical field inside the superconductor is in the $z$ direction, we obtain that
\begin{equation}
\label{Ephi}
E=-\partial_z\phi-{\dot A},
\end{equation}
where $\phi$ is the electrical scalar potential, $\partial_z\phi\equiv \partial\phi/\partial z$ and ${\dot A}\equiv \partial A/\partial t$. The electrical field and the vector potential have zero components in the $x$ and $y$ directions, so that $\partial\phi/\partial x=\partial\phi/\partial y=0$. Then, since for infinitely long conductors $E$ do not depend on $z$, $\partial_z\phi$ is uniform in the whole cross-section. The quantity $\partial_z\phi$ can be calculated taking one point where $E=0$ and using Eq.\ (\ref{Ephi}), obtaining that
\begin{equation}
\label{phiA}
\partial_z\phi(t)=-{\dot A}(x_0,y_0;t),
\end{equation}
where $x_0$ and $y_0$ are the $x$ and $y$ coordinates at some point where $E=0$. For the critical-state model, $E$ always vanishes in the positions where $J=0$. Such position can be either in the current-free core or on the flux fronts (Sec. \ref{s.csm}).

Inserting Eqs. (\ref{Ephi}) and (\ref{phiA}) into Eq. (\ref{Q}) and using ${\dot A}={\dot I}\partial_IA$ and ${\rm d}I={\dot I}{\rm d}t$, we obtain
\begin{eqnarray}
\label{QI}
Q & = & 2\int_{-I_{\rm m}}^{I_{\rm m}} {\rm d}I \int_S J_{\rm ret}(x,y;I) \nonumber\\
& & [\partial_IA(x_0,y_0;I)-\partial_IA(x,y;I)]{\rm d}x{\rm d}y,
\end{eqnarray}
where the current integration is performed in the returning stage. From Eq. (\ref{QI}), we see that $Q$ is independent on the specific $I(t)$ waveform, as long as $I$ increases or decreases monotonically with time in a half cycle. From this feature it is directly deduced that the ac loss due to a sinusoidal current and applied field is independent on their frequency.

The vector potential can be easily calculated from $J$, obtained by means of the numerical procedure described in Secs. \ref{s.minF} and \ref{s.Jcalc}. Then, we calculate $\partial_IA$ at a certain time layer $k$ from the numerically obtained $A$ as
\begin{eqnarray}
\label{dIAnum}
\partial_IA(x,y;I_k) & \approx &\frac{A(x,y;I_{k+1})-A(x,y;I_{k-1})}{I_{k+1}-I_{k-1}},
\end{eqnarray}
where $I_k$ is the current in the time layer $k$. Equation (\ref{dIAnum}) yield much more accurate results of $\partial_IA$ than using finite differences between consecutive time layers. Indeed, according to the mean value theorem, there must exist some current between $I_{k-1}$ and $I_{k+1}$ where the derivative is exactly the right-side part of Eq.\ (\ref{dIAnum}). Equation (\ref{dIAnum}) cannot be used at the boundaries of a half cycle, $I=\pm I_{\rm m}$, since there $\partial_IA$ is not continuous. Therefore, we use finite differences between $k$ and $k+1$ or $k$ and $k-1$ for $I=-I_{\rm m}$ and $I=I_{\rm m}$, respectively.

If in the initial stage the current front monotonically penetrates with increasing $I_{\rm m}$ (or $H_{\rm m}$), it is possible to obtain a formula for the ac loss without derivatives in the vector potential.  This is the case, e.g. of the pure ac transport or the magnetization by an ac external field. For this situation, $J_{\rm rev}$ and $J_{\rm ret}$ obey Eqs.\ (\ref{JrevJin}) and (\ref{JretJin}). Thanks to this feature and following the same deduction as Carr for the pure transport situation \cite{carr04bPhC}, the ac loss from Eq. (\ref{QI}) becomes
\begin{equation}
\label{Qsim}
Q=4J_{\rm c}\int_S s(x,y) \left[{A^k_{\rm m}-A_{\rm m}(x,y)}\right]{\rm d}x{\rm d}y,
\end{equation}
where $A^k_{\rm m}$ and $A_{\rm m}(x,y)$ are those corresponding to the peak values of $I$ and $H_{\rm a}$ and $s(x,y)$ is a function giving the sign of $J_{\rm rev}$. Equation (\ref{Qsim}) with $s=1$ corresponds to that obtained by Norris for the transport case \cite{norris70JPD} and with $s=x/|x|$ it corresponds to the magnetic one given by Rhyner \cite{rhyner02PhC}.

Since the condition of monotonic current front movement is not always valid, it is saver use Eq. (\ref{QI}) for a general treatment of simultaneous ac magnetic field and alternating transport current.


\section{Results and discussion}
\label{s.resdis}

In this section we present our results for the current distribution and the ac loss and we discuss the existing analytical approximations for low and high $b/a$ aspect ratios. We also introduce the dissipation factor $\Gamma=2\pi Q/(\mu_0I_{\rm m}^2)$, characterizing the loss behavior better than the ac loss itself.


\subsection{Current distribution}
\label{s.prof}

In the following we present the current distribution for a rectangular strip with aspect ratio $b/a=0.2$, although the numerical procedure gives accurate results for $b/a$ between 0.001 and 100. We consider several situations of field and current.

First, we study the case of low applied fields. As an example, we plot the current distribution for $I_{\rm m}/I_{\rm c}=0.8$ and $H_{\rm m}/H_{\rm p}=0.08$ in Fig.\ \ref{f.JIp8Hp08}. These profiles are qualitatively similar to those for transport current \cite{aclosrec} with the difference that the current-free core is shifted to the left. In this situation, the current fronts monotonically penetrate from the surface inwards. Thus, the current distribution can be calculated using MEM (Sec. \ref{s.FW}) and the current distribution for the whole cycle can be constructed from that one in the initial stage using Eqs. (\ref{JrevJin}) and (\ref{JretJin}). For this case, the ac loss can be calculated using Eq.\ (\ref{Qsim}), so that the evaluation of $E$ can be skipped.

The most representative situation of combined action of ac field and ac current is that of higher applied fields, such as $I_{\rm m}/I_{\rm c}=0.6$ and $H_{\rm m}/H_{\rm p}=0.72$, presented in Fig.\ \ref{f.JIp6Hp72}. The most significant issue is that the current distribution in the reverse stage is not always a superposition of that for the initial stage. Not even the returning stage is related to the reverse one. However, as can be seen in the figure, the current distribution for $I/I_{\rm m}=1$ (and $H_{\rm a}/H_{\rm m}=1$) in the reverse stage corresponds to that for $I/I_{\rm m}=-1$ (and $H_{\rm a}/H_{\rm m}=-1$) for the returning one, except some numerical deviation. Then, the current distribution for the following reverse stage for a certain transport current $I$ (and applied field $H_{\rm a}$) is the same but with opposite sign with respect to those for the returning stage for transport current $-I$ (and applied field $-H_{\rm a}$), being current distribution periodic in time after the first cycle.

The above specific case (Fig.\ \ref{f.JIp6Hp72}) presents a current-free core, but it is not always the case for higher $H_{\rm m}$ or $I_{\rm m}$. For example, the current-free core is not present for $I_{\rm m}/I_{\rm c}=0.6$ and $H_{\rm m}/H_{\rm a}=1.2$  (Fig.\ \ref{f.JIp6H1p2}), as well as for any case with $I_{\rm m}=I_{\rm c}$, as shown in Fig.\ \ref{f.JI1H2} for $I_{\rm m}=I_{\rm c}$ and $H_{\rm m}/H_{\rm p}=2$. In addition, for all the cases with $I_{\rm m}=I_{\rm c}$ the electromagnetic history is erased at the end of one half cycle (Fig.\ \ref{f.JI1H2}). Thus, for this current amplitude, the returning profiles correspond to the reverse ones with inverted sign of the current density, so that the electromagnetic behavior is simplified. Another issue is that the current distribution for $I_{\rm m}=I_{\rm c}$ have only one boundary between the zone of positive and negative current, while they can have two or more for lower $I_{\rm m}$ (Figs. \ref{f.JIp6Hp72} and \ref{f.JIp6H1p2}).

For other aspect ratios we found the same qualitative behavior as for $b/a=0.2$ described above. As an example, in Fig.\ \ref{f.q5Ip6Hp72} we present the current distribution in the returning curve for $b/a=5$. This corresponds to the situation of the same strip as for Figs. \ref{f.JIp8Hp08} to \ref{f.JI1H2} but with applied field parallel to the wide direction.

\subsubsection{Comparison with analytical limits}
\label{s.cmpcurprof}

It is interesting to compare the sheet current density $K$ in a thin film from Refs. \cite{brandt93PRB,zeldov94PRB}, where it is assumed that current fronts penetrate monotonically, with our results for finite thickness. We calculated $K$ by integrating the current distribution over the thickness. For this situation, in Refs. \cite{brandt93PRB,zeldov94PRB} it is distinguished between the low-field high-current regime, for which all current has the same sign, and the high-field low-current one, when current density with both signs exist. These regimes appear for $I/I_{\rm c}\ge \tanh(H_{\rm a}/H_c)$ and $I/I_{\rm c}\le \tanh(H_{\rm a}/H_c)$, respectively, being $H_c\equiv 2J_{\rm c}b/\pi$. 

In Fig.\ \ref{f.TSprof} we present our numerical calculations of $K$ for the initial stage together with the analytical results for a thin film for $b/a=0.01$, $H_{\rm m}/H_{\rm p}=0.1$ and $I_{\rm m}/I_{\rm c}=1$ (a), belonging to the low-field high-current regime, and $H_{\rm m}/H_{\rm p}=1$ and $I_{\rm m}/I_{\rm c}=1$ (b), as an example for the high-field low-current case. As can be seen in Fig.\ \ref{f.TSprof}(a), for low applied fields all numerical results fall on the analytical curve within the numerical accuracy, while for higher fields, Fig.\ \ref{f.TSprof}(b), there is only coincidence for the profiles corresponding to low current penetration. The discrepancy for higher penetration appears since the assumption of monotonic penetration of current fronts is no longer valid for the analytical solution. As can be seen in Fig.\ \ref{f.JIp6H1p2}(a,b,c), for high $I/I_{\rm c}$ there is a recession of the zone with negative current density in favor of that one with positive current density, being this effect more important for higher $I$. The validity of the analytical solution for thin strips was already discussed in Refs.\ \cite{brandt93PRB,zeldov94PRB}.

We can also compare the numerically obtained current distribution to the analytical solution for a slab in parallel field, for which the current fronts are planar \cite{carr79IEM,zeldov94PRB}. For strips with high $b/a$ in the high-field low-current regime, the calculated current fronts approach to planar ones (Fig.\ \ref{f.q5Ip6Hp72}), the approximation being better for higher $b/a$. This behavior is in agreement to the pure magnetic case \cite{brandt96PRB}. However, for the low-field high-current regime current fronts are similar to that ones for a thin strip with only transport current, which are nonplanar \cite{norris70JPD,aclosrec} and, thus, the slab approximation is no longer valid. 

We have performed numerical simulations for very high applied fields, $H_{\rm m}>5H_{\rm p}$, and have shown that the current fronts approach to vertical planes for any aspect ratio, in accordance to the slab approximation. This is because when the applied field variation is much higher than the field created by the variation of $J$, the first term of ${\cal F}'$ in Eq.\ (\ref{F'def}) can be neglected. Since $A_{\rm a}$ is proportional to $x$, ${\cal F}'$ of the new induced current density is independent of its $y$ location, and the current density profiles must be planar.


\subsection{Total ac loss}

First, we study the ac loss for several $b/a$ aspect ratios and their possible analytical approximations. For this purpose, we present our results of the normalized ac loss $q\equiv 2\pi Q/(\mu_0I_{\rm c}^2)$ as a function of $I_{\rm m}$ and constant $H_{\rm m}$ and as a function of $H_{\rm m}$ and constant $I_{\rm m}$ in Figs. \ref{f.Qp001}(a),\ref{f.Q100}(a),\ref{f.Qp1}(a) and Figs. \ref{f.Qp001}(b),\ref{f.Q100}(b),\ref{f.Qp1}(b), respectively. Figures \ref{f.Qp001}, \ref{f.Q100} and \ref{f.Qp1} are for aspect ratios $b/a=0.001$, 100 and 0.1, respectively. The aspect ratios of $b/a=0.001$ and 0.1 can be used to describe qualitatively YBCO coated conductors and Ag/Bi-2223 tapes, respectively, in perpendicular field. Aspect ratio $b/a=100$ is representative for parallel applied field. For all
figures, we consider $I_{\rm m}$ normalized to $I_{\rm c}$ and $H_{\rm m}$ normalized to the full penetration field $H_{\rm p}$, which for a rectangular strip is \cite{forkl,brandt96PRB}
\begin{equation}
\label{Hp}
H_{\rm p}=J_{\rm c}\frac{b}{\pi}\left[\frac{2a}{b}\arctan\frac{b}{a}+\ln\left(1+\frac{a^2}{b^2}\right)\right].
\end{equation}

The numerical error in the ac loss have been analyzed by using several numbers of elements, current steps and time layers, showing a variation that cannot be appreciated for the axis scale of all figures below.

From Figs. \ref{f.Qp001}, \ref{f.Q100} and \ref{f.Qp1} we see that the ac loss monotonically increases with increasing either the current or the applied field amplitudes. For high applied field, $Q$ increases linearly with $H_{\rm m}$ for constant $I_{\rm m}$ [Figs. \ref{f.Qp001}(b),\ref{f.Q100}(b) and \ref{f.Qp1}(b)]. The ac loss for the low-current limit in figures (a) and the low-applied-field limit in figures (b) is constant, corresponding to the pure transport and pure magnetic case, respectively. This qualitative behavior is consistent with experiments for YBCO coated conductors \cite{ashworth00JAP,ogawa03IES} and Ag/Bi-2223 tapes \cite{rabbers99IES,magnusson99IES,ashworth99aPhC,ashworth00PhC,tonsho03IES,amemiya03aPhC}.

As expected, the loss results for zero applied field and zero transport current are the same to the results for pure transport and pure magnetic situations calculated using MEM in Refs. \cite{aclosrec} and \cite{acxrec}, respectively.

\subsubsection{Analytical limits for the ac loss}
\label{s.cmploss}

In the following we study first the validity of the analytical limits for thin strips and slabs in parallel applied field.

In Fig.\ \ref{f.Qp001} we compare our numerical results of $2\pi Q/(\mu_0I_{\rm c}^2)$ for $b/a=0.001$ (line plus symbols) with the ac loss for an infinitely thin strip calculated by Sch\"onborg \cite{schonborg01JAP} (dash line) from the sheet current distribution obtained in \cite{brandt93PRB} and \cite{zeldov94PRB}. Sch\"onborg expression for $I_{\rm m}=0$ corresponds to the Norris formula for a thin strip with pure transport current \cite{norris70JPD}. The thick continuous line plotted in Fig.\ \ref{f.Qp001} separates the low-field high-current regime from the high-current low-field one, Sec.\ \ref{s.cmpcurprof}. As can be seen in Fig.\ \ref{f.Qp001}(a,b) the ac loss for the low-field high-current regime is well described by the analytical expressions for a thin strip. However, there is a significant deviation for the high-field regime, increasing with increasing field or current. This is so because, as discussed in \cite{brandt93PRB,zeldov94PRB}, the current density formulae for thin strips are only valid for monotonic penetration of current fronts, which appears only for the low-field high-current regime, Fig.\ \ref{f.JIp8Hp08}. The current front penetration deviates more from the monotonic case for higher field and current, so that the formulae for thin strips are less applicable.

In Fig.\ \ref{f.Qp1} we plot our numerically calculated $2\pi Q/(\mu_0I_{\rm c}^2)$ for $b/a=0.1$ (line with symbols) together with that for a thin strip (dash line). In this figure we can see that the thin-film approximation is not valid for $b/a$ for any case except $I_{\rm m}=I_{\rm c}$ and low applied field. 

It is also interesting to compare our numerical results to the formulae for the ac loss obtained by Carr for a slab in parallel applied field assuming planar current fronts \cite{carr79IEM}, which in SI are
{\setlength\arraycolsep{2pt}
\begin{eqnarray}
\frac{2\pi Q}{\mu_0I_{\rm c}^2} & = & \frac{\pi a}{3b}i_{\rm m}^3 \left[1+3\frac{h_{\rm m}^2}{i_{\rm m}^2}\right], \qquad h_{\rm m}\le i_{\rm m} \label{carr1}\\
\frac{2\pi Q}{\mu_0I_{\rm c}^2} & = & \frac{\pi a}{3b}h_{\rm m}^3 \left[1+3\frac{i_{\rm m}^2}{h_{\rm m}^2}\right], \qquad i_{\rm m}<h_{\rm m}\le 1 \label{carr2}\\	
\frac{2\pi Q}{\mu_0I_{\rm c}^2} & = & \frac{\pi a}{b}h_{\rm m}\left[1+\frac{i_{\rm m}^2}{3}\right] 
- \frac{2\pi a}{3b}(1-i_{\rm m})(1+i_{\rm m}+i_{\rm m}^2) \nonumber\\
& & + \frac{2\pi a}{b} i_{\rm m}^2 \frac{(1-i_{\rm m}^2)}{h_{\rm m}-i_{\rm m}} \nonumber\\
& & - \frac{4\pi b}{3a} i_{\rm m}^2\frac{(1-i_{\rm m})^3}{(h_{\rm m}-i_{\rm m})^2}, \qquad h_{\rm m}>1,\label{carr3}
\end{eqnarray}}
where $i_{\rm m}=I_{\rm m}/I_{\rm c}$ and $h_{\rm m}=H_{\rm m}/(J_{\rm c}a)$. The high-field limit of Eq.\ (\ref{carr3}) is 
\begin{equation}
\frac{2\pi Q}{\mu_0I_{\rm c}^2} = \frac{\pi a}{b}h_{\rm m}\left[1+\frac{i_{\rm m}^2}{3}\right], \qquad h_{\rm m}\gg 1. \label{carrhH}
\end{equation}

In Fig.\ \ref{f.Q100} we plot our numerical results of $2\pi Q/(\mu_0I_{\rm c}^2)$ for $b/a=100$ (line with symbols) together with those for a slab calculated from Eqs. (\ref{carr1})-(\ref{carr3}) (dash line). We see that the above formulae for slabs agree well with the numerical results for high fields and low currents, although they do not for low fields and high currents. In Fig.\ \ref{f.Q100} we also see that for $H_{\rm m}$ much above $H_{\rm p}$, the Carr's results approach to the actual loss for any current. These features can be explained from the current distribution, discussed in Sec. \ref{s.cmpcurprof}.

The Carr formula can be also compared to numerical results for any $b/a$. In Figs. \ref{f.Qp001}(a) and \ref{f.Qp1}(a) we include the high-field limit of the ac loss in a slab, Eq.\ (\ref{carrhH}), for the highest values of $H_{\rm m}/H_{\rm p}$ in those graphs (dotted lines). It can be seen that the analytical limit of Eq.\ (\ref{carrhH}) approaches to the numerical results for high $H_{\rm m}$ for $H_{\rm m}/H_{\rm p}\ge 1$ and $H_{\rm m}/H_{\rm p}\ge 5$ for $b/a=0.001$ and $b/a=0.1$, respectively. Numerical calculations for other $b/a$, like $b/a=1$, also agree with Eq.\ (\ref{carrhH}) for high applied field amplitudes. This feature can be explained as follows. For high $H_{\rm m}$, the current fronts are planar, like those for a slab, Sec. \ref{s.cmpcurprof}. Moreover, if $H_{\rm m}$ is high enough, the only relevant contribution to the electrical field, Eqs. (\ref{Ephi}) and (\ref{phiA}), is from the applied vector potential for any aspect ratio. Then, the high-field limit for a slab must be valid for any aspect ratio. Actually, Eq.\ (\ref{carrhH}) can be easily deduced from Eqs. (\ref{Q})-(\ref{phiA}) assuming that ${\dot A}\approx{\dot A}_a=-\mu_0x{\dot H}_a$.


\subsection{Dissipation factor}

Usually ac loss under alternating field and current have been studied as a function of $I_{\rm m}$ and fixing $H_{\rm m}$ or {\it vice versa}, \cite{rabbers99IES,magnusson99IES,ashworth99aPhC,ashworth00PhC,zannella01IES,tonsho03IES,amemiya03aPhC,ashworth00JAP,ogawa03IES,schonborg01JAP}. Here we underline the significance of the ac-loss dependence when increasing simultaneously $I_{\rm m}$ and $H_{\rm m}$ with both parameters proportional to each other along the curve. This situation is found in actual ac devices, such as an alternating magnet. 

As explained below, for $I_{\rm m}\propto H_{\rm m}$ we can see more details of the ac loss behavior if we plot $Q$ normalized to $I_{\rm m}^2$ instead to $I_{\rm c}^2$. Actually, the quantity $2\pi Q/(\mu_0I_{\rm m}^2)\equiv \Gamma$ is proportional to the ac loss of a winding per the stored magnetic energy averaged during the cycle duration. Thus, $\Gamma$ can be regarded as a dissipation factor. Moreover, $\Gamma$ for only transport current is proportional to the imaginary part of the self-inductance, defined in \cite{gomory98aPhC}, and for only applied magnetic field $\Gamma$ is proportional to the imaginary part of the ac susceptibility \cite{chen91JAP,gomory97SST}.

In Fig.\ \ref{f.Qi2p1} we present our numerical results of $\Gamma$ for $b/a=0.1$ as a function of $I_{\rm m}$ when $H_{\rm m}$ is varied proportionally to $I_{\rm m}$ as $H_{\rm m}/H_{\rm p}=\alpha I_{\rm m}/I_{\rm c}$, where $\alpha$ is a constant (line with symbols). This figure shows that for the low $I_{\rm m}$ (and $H_{\rm m}$) limit, $\Gamma$ increases proportionally to $I_{\rm m}$ (or $H_{\rm m}$), which corresponds to a dependence proportional to $I_{\rm m}^3$ for the ac loss. Moreover, for high $\alpha$, $\Gamma$ decreases with increasing $I_{\rm m}$ with a slope in log-log scale slightly higher than $-1$ (and a slope around $1$ for the ac loss), presenting a peak at a certain value of $I_{\rm m}$ (or $H_{\rm m}$). We notice that in a log-log plot of $q$ against $I_{\rm m}$, the ac loss always increase with $I_{\rm m}$, appearing curves very similar to straight lines with a slight change in the slope. However, for $\Gamma$ the qualitative behavior of the loss with varying $I_{\rm m}$ and $\alpha$ is more evident.

The linear dependence of the dissipation factor with $I_{\rm m}$ (and $H_{\rm m}$) for the low-field limit is characteristic of the critical-state model, also found for the pure transport and pure magnetic situations \cite{aclosrec,acxrec}. This is so since for low enough $I_{\rm m}$ the current distribution is approximately parallel to the surface, as well as the magnetic field in the region with nonzero current density \cite{acxrec}. This means that the loss factor will increase as $I_{\rm m}$ (and $Q$ as $I_{\rm m}^3$) at low levels of excitation in a superconductor winding of any shape or number of turns. However, for strips with very small $b/a$, such as $b/a=0.001$, the linear dependence of $\Gamma$ with $I_{\rm m}$ appears only for very small $I_{\rm m}$, presenting for higher $I_{\rm m}$ the $I_{\rm m}^2$ dependence typical for thin films \cite{acxrec}.

For comparison, in Fig.\ \ref{f.Qi2p1} we also include the loss factor for only transport current (dash-dot line) and that for only applied field (dash lines), extracted from the tables in \cite{aclosrec} and \cite{acxrec}, respectively, and interpolating for intermediate values of $I_m$ or $H_m$ when needed. In a $\Gamma (I_{\rm m}/I_{\rm c})$ representation, the curve for the pure magnetic ac loss is different for each $\alpha$. We see that for low $\alpha$, $\Gamma$ approaches to that one for only transport for any $I_{\rm m}$ and for high $\alpha$, $\Gamma$ approaches to that one for only magnetic field for not very high $I_{\rm m}$. 

For a finer approximation for high $\alpha$, we can consider the following loss factor
\begin{equation}
\Gamma(I_{\rm m},H_{\rm m})\approx\frac{2\pi}{\mu_0I_{\rm m}^2}\left[Q_{\alpha\rightarrow\infty}(I_{\rm m},H_{\rm m})\right],\label{Ghal}
\end{equation}
where $Q_{\alpha\rightarrow\infty}$ is an approximated ac loss as
\begin{equation}
Q_{\alpha\rightarrow\infty}(I_{\rm m},H_{\rm m})\equiv Q(I_{\rm m}=0,H_{\rm m})+\frac{2a\mu_0H_{\rm m}I_{\rm m}^2}{3I_{\rm c}}
\label{Qhal}.
\end{equation}
The first term of Eq. (\ref{Qhal}) is the ac loss with only applied magnetic field, while the second one is the high-field limit for an slab, Eq.\ (\ref{carrhH}), subtracting the ac loss for $I_{\rm m}=0$. In Fig.\ \ref{f.Qi2p1} we plot the loss factor of Eq.\ (\ref{Ghal}) for $\alpha=2$ and 5, calculated from the tables of numerically calculated ac susceptibility in \cite{acxrec} and Eq.\ (\ref{Ghal}). We see that the approximation of Eq.\ (\ref{Ghal}) improves with increasing $\alpha$, being almost overlapped with our numerical results for $\alpha\le5$.

For low $b/a$, such as $b/a=0.001$ or lower, the ac loss for $\alpha=0$ approaches to the Norris formula for thin strips \cite{norris70JPD}, if $I_{\rm m}$ is not very low \cite{aclosrec}. For the high-$\alpha$ limit we can obtain an analytical solution of $\Gamma$ by inserting the formula for the ac loss in a thin strip with $I_{\rm m}=0$, Ref.\ \cite{brandt93PRB}, into Eq.\ (\ref{Ghal}), obtaining
\begin{equation}
\label{GhalTS}
\Gamma=\frac{2H_{\rm m}}{H_c}\left[{\frac{1}{3}+\frac{I_{\rm c}^2}{I_{\rm m}^2}\left({\frac{2H_c}{H_{\rm m}}\ln\cosh\frac{H_{\rm m}}{H_c}-\tanh\frac{H_{\rm m}}{H_c}}\right)}\right],
\end{equation}
where $H_c=2bJ_{\rm c}/\pi$. Equation (\ref{GhalTS}) is not valid for very low $H_{\rm m}/H_c$, since the thin strip approximation for only applied field is not valid for this range \cite{acxrec}.
For intermediate $\alpha$, $\Gamma$ can be approximated from the Sch\"onborg's formula for the ac loss in thin strips \cite{schonborg01JAP}, as long as $I_{\rm m}\ge \tanh[\pi H_{\rm m}/(2aJ_{\rm c})]$, Sec. \ref{s.cmploss}.


\section{Comparison with experiments}
\label{s.exp}

The results of our ac loss calculations presented in Figs.\ \ref{f.Qp001}, \ref{f.Q100}, and \ref{f.Qp1} qualitatively agree with published measurements for Ag/Bi-2223 tapes \cite{rabbers99IES,magnusson99IES,ashworth99aPhC, ashworth00PhC,tonsho03IES,zannella01IES,amemiya03aPhC} and YBCO coated conductors \cite{ashworth00JAP,ogawa03IES}. It is interesting to analyze in detail Fig. 10 of Ref. \cite{ashworth00JAP}. There, it is  shown a comparison between the measured ac loss in a YBCO coated conductor and the theoretical one for a thin strip in the critical state model, evaluated from the current distribution in \cite{brandt93PRB,zeldov94PRB}. It can be seen that the measured ac loss lay below the thin strip approximation, in agreement with our numerical results of Fig. \ref{f.Qp001}. As discussed in Sec.\ \ref{s.cmploss}, this is so because the thin strip calculations in Refs.\ \cite{ashworth00JAP,schonborg01JAP} are not valid for high applied fields. This shows that our numerical calculations can be used to simulate the ac loss in YBCO coated conductors.

In order to perform a more detailed comparison, we measured the loss factor for a commercial Ag/Bi-2223 tape with 37 filaments manufactured by Australian Superconductor. The sample was of 8cm length and 3.2$\times$0.31mm cross-section with a critical current of 38A in self field at 77K. The superconducting core cross-section was roughly elliptical with dimensions $2a\times 2b=3.0\times 0.13$mm. The measurements were performed at a 72Hz frequency and 77K temperature; the details of the experimental technique will be presented elsewhere. We present the measured results in Fig.\ \ref{f.meas} (dot line with symbols) together with numerical calculations for a rectangular strip with the same thickness, width and critical current (solid lines). We notice that for the theoretical curves we do not fit any parameter to the measured ones. In Fig.\ \ref{f.meas} we label the curves with the parameter $2aH_{\rm m}/I_{\rm m}$ instead of $\alpha$ in order to avoid supposing any model for performing the measurements. For comparison, we also include the loss factor assuming elliptical cross section for only transport current \cite{norris70JPD} (lower dash curve) and only applied field for the highest $2aH_{\rm m}/I_{\rm m}$ \cite{acxell} (upper dash curve).

From Fig.\ \ref{f.meas} we see that the main qualitative features of the measurements correspond to the behavior of a strip assuming the critical state model, except close to $I_c$ for high $2aH_{\rm m}/I_{\rm m}$. This can be explained from the magnetic field $B$ dependence of $J_c$, for which $J_c$ decreases with increasing $|B|$. Then, for higher $H_{\rm m}$, $I_{\rm c}$ is lower and a normal resistive current appears in the silver for $I_{\rm m}<I_{\rm c}$, adding a certain contribution to $\Gamma$. Figure \ref{f.meas} shows that there is a better agreement between the measured $\Gamma$ and that for an elliptical bar in the critical state model approximation than for the rectangular one, explained by the overall shape of the tape superconducting core. Moreover, the fact that the multifilamentary superconducting core behaves as a single solid wire for any $H_{\rm m}$
suggests that the interfilamentary currents in the tape are saturated due to the high length of the sample \cite{fukumoto95APL}.


\section{Conclusions}
\label{s.concl}

In this paper we have presented a systematic theoretical study of the current distribution and ac loss in a rectangular strip transporting an alternating transport current $I$ in phase with an applied field ${\bf H}_{\rm a}$ perpendicular to the current flow. We assumed that the superconductor follows the critical state model with a constant $J_{\rm c}$. With this assumption, we have developed a numerical procedure which takes into account the finite thickness of the strip. General features of the critical state model in such circumstances have been discussed. In order to do a systematic study, we have performed extensive numerical calculations for several aspect ratios and current and applied field amplitudes, $H_{\rm m}$ and $I_{\rm m}$ respectively.  Finally, we have performed measurements on Ag/Bi-2223 tapes to be confronted with calculations. Good qualitative and quantitative agreement without fitting parameters have been found.

The results for the current distribution have shown a rich phenomenology due to the metastable nature of the electrical currents flowing in the superconductor. For low $H_{\rm a}$ and high $I$, the current distribution is qualitatively similar to the pure transport situation. Then, $J$ at the reverse and returning stages are a superposition of $J$ in the initial one [Eqs. (\ref{JrevJin}) and (\ref{JretJin})]. However, it is not the same for high $H_{\rm a}$ or low $I$ due to the nonmonotonic penetration of current fronts. In general, the returning stage can be deduced from that one in the first reverse stage. Only after the first cycle, the behavior becomes periodic.

The ac loss $Q$ has been accurately calculated for the thickness-to-width aspect ratios, $b/a$=0.001, 0.1 and 100, in order to qualitatively describe YBCO coated conductors and Ag/Bi-2223 tapes with applied fields in the transverse direction ($b/a=$0.001 and 0.1, respectively) and in the parallel one ($b/a=$100). Their current and applied field dependence is in accordance with published measured data for YBCO coated conductors \cite{ashworth00JAP,ogawa03IES} and Ag/Bi-2223 tapes \cite{rabbers99IES,magnusson99IES,ashworth99aPhC,ashworth00PhC,tonsho03IES,amemiya03aPhC}. We have shown that the ac loss behavior can be better characterized by means of the loss factor $\Gamma=2\pi Q/(\mu_0I_{\rm m}^2)$ studied as a function of $I_{\rm m}$ with $H_{\rm m}$ proportional to $I_{\rm m}$. We have measured the loss factor in actual Ag/Bi-2223 tapes, obtaining a good agreement with the calculations.

We have also presented a detailed study of the analytical limits for $Q$ and $\Gamma$ and their applicability. For thin samples like YBCO coated conductors, $b/a\le 0.001$, the current profiles and the ac loss only approach to those for the analytical limit for thin strips \cite{schonborg01JAP} for the low-field high-current regime only. The thin film approximation is never valid for $b/a \sim 10$, such as for Ag/Bi-2223 tapes. We have also studied the slab limit, obtaining that for the situation of parallel field, $b/a=100$, the slab approximation is not valid for the transport-like regime (low-$H_{\rm m}$ and high-$I_{\rm m}$). However, the high-field limit for the slab approximation can be used for any aspect ratio provided that $H_{\rm m}$ is high enough.


\section*{Acknowledgements}

We acknowledge M. Vojen\v ciak for valuable technical support in the measurements and D.-X. Chen and A. Sanchez for comments in the writting of the paper. ASTRA interchange project is acknowledged for financial support.



\newpage
\null

\begin{figure}[h]
\caption{Sketch of the tape cross-section and division into elements for the calculations. The transport current $I$ and the applied field ${\bf H}_{\rm a}$ are directed in the positive $z$ and $y$ directions, respectively.}\label{f.sketch}
\end{figure}

\begin{figure}[h]
\caption{Current distribution in the initial stage for the low-field and high-current regime. Specific parameters are $b/a=0.2$, $H_{\rm m}/H_{\rm p}=0.08$, $I_{\rm m}/I_{\rm c}=0.8$ and $I/I_{\rm m}=H_{\rm a}/H_{\rm p}=0.2$ (a), 0.6 (b) and 1 (c). Local current density is $+J_{\rm c}$ for the black region and zero for the white one.}\label{f.JIp8Hp08}
\end{figure}

\begin{figure}[h]
\caption{Current distribution for $b/a=0.2$, $H_{\rm m}/H_{\rm p}=0.72$, and $I_{\rm m}/I_{\rm c}=0.6$ at several instants of the ac cycle. Figures (a,b,c) are for the initial stage with $I/I_{\rm m}=0.2$, 0.6, and 1, respectively, figures (d,e,f) are for the reverse stage with $I/I_{\rm m}=0.6$, $-0.2$ and $-1$, respectively, and figures (g,h,i) are for the returning stage with $I/I_{\rm m}=-1$, 0.2, and 1, respectively. In the black regions the local curent density is $+J_{\rm c}$, in the light gray zones it is $-J_{\rm c}$, and in the white ones it is zero.}\label{f.JIp6Hp72}
\end{figure}

\begin{figure}[h]
\caption{Current distribution for $b/a=0.2$, $H_{\rm m}/H_{\rm p}=1.2$ and $I_{\rm m}/I_{\rm c}=0.6$ at several instants of the ac cycle. Figures (a,b,c) are for the initial stage with $I/I_{\rm m}=0.2$, 0.6, and 1, respectively, and figures (d,e,f) are for the reverse stage with $I/I_{\rm m}=0.6$, $-0.2$ and $-1$, respectively. In the black regions the local curent density is $+J_{\rm c}$ and it is $-J_{\rm c}$ in the gray zones.}\label{f.JIp6H1p2}
\end{figure}

\begin{figure}[h]
\caption{Current distribution at the reverse stage for $b/a=0.2$, $H_{\rm m}/H_{\rm p}=2$, $I_{\rm m}/I_{\rm c}=1$ and $I/I_{\rm m}=0.6$ (a), $-0.2$ (b), and $-1$ (c), respectively. In the black regions the local curent density is $+J_{\rm c}$ and it is $-J_{\rm c}$ in the gray zones.}\label{f.JI1H2}
\end{figure}

\begin{figure}[h]
\caption{Current distribution at the reverse stage for $b/a=5$, $H_{\rm m}/H_{\rm p}=0.72$, $I_{\rm m}/I_{\rm c}=0.6$ and $I/I_{\rm m}=0.6$ (a), $-0.2$ (b), and $-1$ (c), respectively. In the black regions the local curent density is $+J_{\rm c}$ and it is $-J_{\rm c}$ in the gray zones.}\label{f.q5Ip6Hp72}
\end{figure}

\begin{figure}[h]
\caption{Sheed current density $K$ in the initial stage as a function of $x$ for $b/a=0.01$, $I_{\rm m}=I_{\rm c}$, and $H_{\rm m}/H_{\rm p}=0.1$ (a) and 1 (b) at several instantaneous $I$ (and $H_{\rm a}$). Lines are for the thin strip limit from Refs. \cite{brandt93PRB,zeldov94PRB} and symbols are for our numerical calculations. For the numerical results, $K$ is the integral of $J$ over the sample thickess.}\label{f.TSprof}
\end{figure}

\begin{figure}[h]
\caption{Normalized ac loss $2\pi Q/(\mu_0I_{\rm c}^2)$ for $b/a=0.001$ as a fuction of $I_{\rm m}/I_{\rm c}$ for several $H_{\rm m}/H_{\rm p}$ (a) and as a fuction of $H_{\rm m}/H_{\rm p}$ for several $I_{\rm m}/I_{\rm c}$ (b). Lines with symbols are for our numerically calulated results, dash lines are for the thin strip limit from \cite{schonborg01JAP}, the thick solid line separates the low-field and high-current regime from the high-field one in a thin strip \cite{schonborg01JAP}, and dotted lines (for $H_{\rm m}/H_{\rm p}$=1 and 2) correspond to the high-field limit for slabs [Eq. (\ref{carrhH})].}\label{f.Qp001}
\end{figure}

\begin{figure}[h]
\caption{Normalized ac loss $2\pi Q/(\mu_0I_{\rm c}^2)$ for $b/a=100$ as a function of $I_{\rm m}/I_{\rm c}$ (a) and as a fuction of $H_{\rm m}/H_{\rm p}$ (b). Solid lines with symbols are our numerically calulated results and dash lines are for the slab approximation from Eqs. (\ref{carr1})-(\ref{carr2}) \cite{carr79IEM}.}\label{f.Q100}
\end{figure}

\begin{figure}[h]
\caption{Normalized ac loss $2\pi Q/(\mu_0I_{\rm c}^2)$ for $b/a=0.1$ as a fuction of $I_{\rm m}/I_{\rm c}$ (a) and as a fuction of $H_{\rm m}/H_{\rm p}$ (b). Lines with symbols are for our numerically calulated results, dash lines are for the thin strip limit from \cite{schonborg01JAP}, and dotted lines (for $H_{\rm m}/H_{\rm p}$=2 and 5) correspond to the high-field limit for slabs [Eq. (\ref{carrhH})].}\label{f.Qp1}
\end{figure}

\begin{figure}[h]
\caption{Loss factor $\Gamma\equiv 2\pi Q/(\mu_0I_{\rm m}^2)$ as a function of $I_{\rm m}/I_{\rm c}$ with $H_{\rm m}$ proportional to $I_{\rm m}$ as $H_{\rm m}/H_{\rm p}=\alpha I_{\rm m}/I_{\rm c}$ for several $\alpha$. Solid lines with symbols are for our numerical calculations, the dash-dot line corresponds to only transport current, the dash ones are for only applied magnetic field, and dotted lines correspond to the high-$\alpha$ approximation from Eqs. (\ref{Ghal}) and (\ref{Qhal}).}\label{f.Qi2p1}
\end{figure}

\begin{figure}[h]
\caption{ Calculated loss factor $\Gamma$ together with experimental data from a commercial Ag/Bi-2223 tape. Dot lines with symbols are for measurements, solid lines correspond to numerical calculations assuming a rectangular cross-section, and the two dash ones are for elliptical cross-section for $H_{\rm m}=0$ (lower curve) and $I_{\rm m}=0$ (top curve) from Refs. \cite{norris70JPD} and \cite{acxell}, respectively.}\label{f.meas}
\end{figure}

\end{document}